\documentstyle[preprint,prc,aps]{revtex}
\begin{document}
\preprint{DOE/ER/40561-141-INT94-00-5, {\it revised version}}
\title{Source Dimensions in Ultrarelativistic Heavy Ion Collisions}
\author{
M. Herrmann$^{(1)}$\thanks{Internet
 address: herrmann@phys.washington.edu},
G.F. Bertsch$^{(1,2)}$\thanks{Internet
 address: bertsch@phys.washington.edu}}
\address{
$^{(1)}$ Institute for Nuclear Theory, HN-12, University of
Washington, Seattle, WA  98195\\
$^{(2)}$ Physics Department, FM-15, University of Washington,
Seattle, WA 98195}
\date{November 18, 1994}
\maketitle
\begin{abstract}
Recent experiments on pion correlations, interpreted as
interferometric measurements of the collision zone, are compared
with  models that distinguish a prehadronic phase and a
hadronic phase.  The models include prehadronic longitudinal
expansion, conversion to
hadrons in local kinetic equilibrium, and rescattering of the
produced hadrons. We find that the longitudinal and outward radii are
surprisingly sensitive to the algorithm used for two-body collisions.
  The longitudinal radius measured in collisions
of 200 GeV/u sulfur nuclei on a heavy target requires the
existence of a prehadronic phase which converts to the hadronic
phase at densities around 0.8-1.0 GeV/fm$^3$.  The transverse
radii cannot be reproduced without introducing more complex
dynamics into the transverse expansion.
\end{abstract}

\narrowtext
\section{Introduction}\label{sec:intro}
Now that a second generation of nucleus-nucleus collision
experiments has been completed at the CERN SPS \cite{qm93}, one may
ask what has been learned about the dynamics of matter at
high density.  In particular, we want
to examine the measurements of interferometric pion correlations,
also known as Hanbury-Brown--Twiss (HBT) correlations, by the two
collaborations NA35 \cite{NA35:be} and NA44 \cite{NA44:be}.
 In principle such experiments
can provide quite detailed information about the space-time
history of the collision\cite{Bo90}. Furthermore the existence
of a long-lived source would show as an unambiguous signal
in the measurement\cite{Pr86,Be89}.

Correlations are often analyzed to give source radii,
and the measured numbers should impose some constraints on the
possible dynamics.  A theoretical model can be tested by
modeling the collision
process, and comparing model predictions with the experimental
results.  One difficulty in drawing firm conclusions is that
there is a great deal of freedom in the construction of these
models.  Another problem, as we will see, is that the correlation
shape is not well described by a single number, the ``source
size'', and so it is better to compare models with the
correlation function actually measured experimentally.

Our objective is to consider a broad class of models that do not
go beyond established or at least widely-accepted physics, to see
whether the experimental results point to new physics.  This
work is a continuation of a study by Welke, et al.\cite{Wel94},
which considered earlier data and had much more detailed model assumptions.
Before proceeding, we have two remarks.  The first is that the
two experiments, NA35 and NA44, are hardly consistent with each
other in the values reported for source radii.  Before any
general conclusions can be drawn from the measurements, it will
be necessary to resolve the disagreement.  We also note that
there was recently reported\cite{Su93,Su94} a study with a model, based
on conventional physics, and good agreement was found with
available data.  Our conclusion is opposite: we shall find that
models
%including only conventional physics
without explicit transverse expansion in the prehadronic phase
cannot explain the body of the data.

Our paper is organized as follows.  In Sec.\ \ref{sec:mod} we discuss the
physical ingredients of conventional models and develop a
general
parameterization of the source function.  In Sec.\ \ref{sec:cf} we discuss
the Gaussian source size and its limitations.  In Sec.\ \ref{sec:res}  we
analyze what constraints data place on the models.
Section \ref{sec:con} summarizes our study and gives an outlook.

\section{Models and parameterizations}\label{sec:mod}
We model the evolution of the system in two stages,
a prehadronic, high density phase\footnote{We
use the term `phase' just to distinguish the stages, not to
imply a thermodynamic phase transition.}
 followed
by a hadronic gas phase.  Many aspects of this two-phase dynamics
are subject to parameterization, but we feel we can explore the
parameter space sufficiently that one is testing a small set of
basic assumptions about the dynamics.  These assumptions are:\\
1) the prehadronic phase only expands longitudinally;
2) the conversion to the hadronic phase is smooth and produces
hadrons with a locally thermalized kinetic distribution.  \\
We now discuss the details of the modeling.

\subsection{High density phase}
This is the most interesting, but least understood part of the
evolution.  However, we do not need a great deal of information
about the structure of this phase; a knowledge of the energy and
momentum densities at the hadronization time should be sufficient.
To be modeled are the transverse and longitudinal distributions
of these quantities.

It is safe to assume that the initial transverse energy distribution
follows the overlap density of the collision partners.  In the
detailed comparisons we will only look at central collisions of
sulfur on heavy targets, so we make take the distribution to
follow the transverse density of sulfur.  We determined this from
the charge density of Ref. \cite{VJV87}, integrating the radial density
with respect to longitudinal distance.  This integral is fit
very well by the following function, which we used in further
numerical modeling
$$  \rho_\perp(r_\perp) ={dn\over dr_\perp^2} = a (1+c r_\perp^2+dr_\perp^4)
\exp(-r_\perp^2/b).$$
The parameters have the following values:  $a = 1.791$ fm$^{-3}$;
$b = 3.79$ fm$^2$; $c = 0.196$ fm$^{-2}$; $d = 0.021$ fm$^{-4}$.
The r.m.s. radius of
$^{32}$S is 3.2 fm \cite{VJV87}, and the r.m.s. transverse radius is
$\sqrt{\langle r_\perp^2 \rangle} = \sqrt{2/3}
\sqrt{\langle r^2\rangle} \approx 2.6$ fm.

We also assume that there is no transverse expansion during the
prehadronic phase.  Certainly there is no initial radial
velocity, but it is not clear that it can be neglected completely
if the prehadronic phase lasts a long time.

In the longitudinal direction, there is a rapid expansion that
must be built into the models.  Two distinct pictures are
possible for how this happens.  One kind of model\cite{Ko86} is based on the
QCD-inspired picture of particle production by string breaking.
The string is produced at a point in space-time and expands
longitudinally.
Particles are produced over the full initial range of
longitudinal rapidities $y$, with the time and longitudinal
position $(t,z)$ of the
production point characterized by an equal spatial rapidity
$$\eta = \mbox{\rm tanh}^{-1}(z/t) \approx y.$$
Another possibility is the Landau picture, in which produced
matter starts out at rest, with a small but finite
longitudinal extension.  Here the high longitudinal momenta arise
from the hydrodynamic expansion.  Models interpolating between
these possibilities have been considered in Ref. \cite{Ve94}
with
the result that the pure Landau picture gives single-particle
rapidity distributions that are too narrow.  We will insist that
the average rapidity of the produced particles be given by their
spatial rapidity, so the distribution of produced particles will
have the form\footnote{By assuming this form, we neglect the
finite thickness of the Lorentz-contracted projectile and target
nucleus.
We have also tried distributions smeared out along the beam axis due to
the target thickness, and found no appreciable effect on the results
reported in Sec. \protect{\ref{sec:res}}.}
\begin{equation}\label{eq:1}
\frac{d^6 N}{d \eta\ dr_{\perp}^2\ dy\ dm_{\perp}^2}
={\cal N}f_\eta(\eta)f_y(y-\eta,p_\perp)\rho_\perp(r_\perp).
\end{equation}

We have little a priori guidance on the the function
$f_\eta$ in Eq.\ (\ref{eq:1}), and we consider it to be completely open
to parameterization.  However, the predicted single-particle
distribution will depend on $f_\eta$, and so it is actually
strongly constrained by the data.  We may therefore assume
some simple form; we take the Gaussian,
$$ f_\eta = \exp(-\eta^2/2(\Delta\eta)^2), $$
which has as a parameter $\Delta \eta$, the r.m.s. dispersion in $\eta$.

To interpret the meaning of the parameter $\Delta \eta$ we note that  for
$\Delta \eta = \infty$ the initial distribution is boost invariant,
i.e., corresponds to the ideal Bjorken scenario\cite{Bj83}. For
$\Delta \eta = 0$ on the other hand we get a initial distribution
that is consistent with the Landau picture of a source that is fully
stopped and expands starting at $\eta = 0$. The parameter $\Delta \eta$
therefore interpolates between the two pictures\cite{Ve94}.

\subsection{Formation of the hadronic phase}
We assume that
the conversion to hadronic matter is controlled only by the
local density as determined by the time evolution of Eq.\ (\ref{eq:1}).
This excludes the possibility of inhomogeneities such as
quark-gluon droplets.  The density at which the conversion takes
place, $n_h$, is an important parameter of the model.
Because of the spatial inhomogeneity in the transverse direction,
the conversion will not take place at the same proper time.  Note
that even with an instantaneous source in proper time, there
would be a range of times in any particular reference frame.
In any case, it will be useful to infer the average proper time
for the formation of the hadronic gas.

Another important question at the conversion point is the composition
and kinetic distribution of the newly-formed hadronic gas.  One
obvious possibility is a chemical and kinetic local equilibrium.
Another possibility is that the distribution carries over from
the distribution formed in hadron-hadron collisions.

We will assume in all our modeling that the kinetic distribution
of particles is thermal with respect to the local frame defined
by $\eta$.  This is certainly plausible, as it will turn out
that the prehadronic phase lasts a considerable time compared to
estimates of the kinetic equilibration time.
The source function $f_y$ will thus be taken to have a thermal form,
$$ f_y = {m_{\perp} \cosh \left(y-\eta\right)\over\left(\exp\left(
m_{\perp} \cosh \left(y-\eta\right)/T\right)-1\right)}. $$
This has as a parameter the kinetic temperature $T$.
The transverse momentum spectra from $pp$ collisions suggest a
temperature of $T=130$ MeV.  However, we demand of our model
that it fit the heavy ion data, and this will require
a slightly higher value.

The chemical composition of the newly-produced hadronic matter
is a much more difficult question, because there is no direct
information about the composition in heavy ion collisions.  For
$pp$ collisions the ratio of rho mesons to final state pions has been
measured\cite{pp1,pp2,pp3}, and is in the range 0.09-0.13.
Note that this is significantly smaller than for the hadronic
jets produced in $e^+-e^-$ annihilation.
Our model also includes omega mesons.
The omega mesons have an important effect on the pion source at
large distance\cite{Pa92}.
Since the
main difference between rho and omega mesons is their isospin,
it is reasonable to suppose that they are produced as the ratio
of isospin degeneracies, $N_\rho : N_\omega = 3:1$.  Assuming
that all pions are produced either directly or via rho and omega
decay, this translates into a ratio of rho mesons to pions at
hadronization time of $w_{\rho/\pi}=0.12-0.21$.  Other authors have used
thermal parameterizations of the particles abundances\cite{Sc93}
which lead to a value near 0.10 for the latter ratio.  We will
test the sensitivity of the extracted source size to this ratio.

We do not explicitly include
$\eta$-mesons in our simulation.  From HELIOS data \cite{HE92}
we estimate that the $\eta$'s contribute 5\% to the total number
of pions in the final state. Due to their {\it very} long lifetime
their contribution will only reduce the intercept of the correlation
function by 0.1 and will therefore have no influence on the determination
of the source radii.

A further source of secondary pions is the decay of the $K^*$ resonance.
In pp-collisions at 400 GeV the number of produced  $K^*$ resonances
was found to be similar to the number of produced $\omega$'s \cite{pp3}.
As there is only one pion produced per $K^*$ decay we neglect this
contribution.

Recently a considerable net proton number $dN_p/dy\approx 8$ has
been reported for S + Ag and S + Au central collisions \cite{NA35:had}.
However, the number of mesons is more than an order of magnitude
larger, and we think the presence of the baryons can be safely
ignored for observables such as the HBT radii.

Finally we ignore Bose symmetry in the hadronic dynamics. It has been
shown that this can significantly alter single-particle observables
when the initial phase-space density is high. However there seems to
be no good quantitative method for dealing with it, and up to now
there is no experimental evidence for strong Bose effects.

\subsection{Hadronic gas}
As discussed in the previous section, the hadronic gas phase
consists of pi, rho and omega mesons. These particles are
allowed
to propagate between interactions, to scatter pairwise, and to
decay. This is a dilute gas approximation, which of course
breaks down at high densities
either because multiparticle interactions become important or
because the mesons are no longer the appropriate
degrees of freedom.  In either case, we may view the parameter $n_h$
as determining the density at which the dilute hadronic gas picture becomes
valid.

The most important (and best known) part of the elastic
hadron-hadron cross-sections is the interaction among the
pions. We use the momentum-dependent, isospin-averaged $\pi \pi$
cross section calculated from phase-shifts \cite{BG88}.
The cross sections for other
mesons are not known.  Given that the pi-pi cross section is
around 20 mb, it is likely that the other cross sections are
within the range of 10-40 mb, and we consider this as the
allowable parameter range.
Inelastic cross sections are more problematic, except for the
well-known $\pi\pi\rightarrow \rho$ cross section.  In principle,
explicit inclusion of this process should lead to larger source
sizes because the rho meson will propagate some distance before
decaying.  However, in the actual modeling it makes no
appreciable
change in the HBT radii whether or not this process is included.
Other inelastic processes are neglected.
It seems likely that
pion-number changing processes have too small a rate to affect
the dynamics on the time scale of the collision\cite{Go93}.

\section{Correlation function and radii}\label{sec:cf}
We now summarize some of the basic formulas for the
interferometric analysis of correlations.  First of all, the
correlation function is defined as the ratio of the two-particle
probability
$P_2$  to the product of the single-particle probabilities $P_1$,
\begin{equation} \label{eq:cdefprob}
C\left(\vec{p_1},\vec{p_2}\right) \equiv \frac{\displaystyle
P_2\left(\vec{p_1},\vec{p_2}\right)}{\displaystyle
P_1\left(\vec{p_1}\right) P_1\left(\vec{p_2}\right)}.
\end{equation}
It is convenient to replace the single-particle momentum variables
$\vec{p_1}$ and $\vec{p_2}$ with the
average momentum $\vec{Q}$
 and relative  momentum $\vec{q}$
\begin{eqnarray} \label{eq:Qqdef}
\vec{Q} & \equiv & \left(\vec{p_1} + \vec{p_2}\right) / 2, \\
\vec{q} &\equiv& \left(\vec{p_1} - \vec{p_2}\right),\nonumber
\end{eqnarray}
and we shall use these variables as the arguments of the
correlation functions below.
Making certain statistical assumptions\cite{We94} about the
production of particles, the correlation function can be calculated from
the single-particle source function $g(x,p)$ by the
formula\cite{Pr84,We94}
\begin{eqnarray}\label{eq:cdefwig}
C\left(\vec{Q},\vec{q}\right) \equiv 1 +
\frac{\displaystyle
\int d^4 x\ \ \int d^4 y\ \ g\left(x, Q\right) g\left(y, Q\right)
\cos\left(q (x-y)\right)}{
\displaystyle
 \int d^4 x\ \ g\left(x, Q+q/2\right) \int d^4 y\ \ g\left(y, Q-q/2\right)
}.
\end{eqnarray}
Here the variables $Q$ and $q$ are four-vectors with time-like
components given by  $Q_0=
\frac{1}{2} (\omega(\vec{Q}+\vec{q}/2) +
\omega(\vec{Q}-\vec{q}/2)) $
and $q_0 =
\omega(\vec{Q}+\vec{q}/2) - \omega(\vec{Q}-\vec{q}/2) $.
Note that the four-momentum argument of the source function in
the numerator is off-shell, i.e. $Q^2 \neq m_{\pi}^2$.

One often discusses correlations for fixed $\vec{Q}$, in which
case the $\vec{q}$ dependence leaves three degrees of freedom
in the correlation function.  We single out three orthogonal
axes to vary $\vec{q}$, working in a frame in which the component
of $\vec{Q}$ along the beam axis vanishes.  The three components
of the correlation function are: \\
1. the longitudinal correlation function $C_l$ with $\vec{q}$
parallel to the beam axis,  $\vec{q} || \hat{z}$;\\
2. the outward correlation function $C_o$ is defined with
$\vec{q}$  tranverse to the beam axis and along $\vec{Q}$,
$\vec{q} || \vec{Q}$;\\
3.  the sideward correlation function $C_s$ with
$\vec{q}\perp\vec{Q}$
and $\vec{q}\perp\hat{z}$.

It is convenient to characterize each of these correlation
functions by a single number, the source radius.  This is
usually taken as the r.m.s. extension of the source along some
axis, but let us start with an operational
definition in terms of correlation functions themselves.  We
define the radii by
\begin{eqnarray}\label{eq:raddef}
R_{i}^2 \equiv - \vec{\nabla}^2_{q_i} C(\vec{Q},\vec{q})_{q=0},
\end{eqnarray}
where $i$ labels the three possible directions.
This is double the mean square source size defined by
$\langle r_i^2 \rangle$,
if the single-particle source has no
momentum dependence.  In general, the relation is\cite{We94}
\begin{eqnarray}\label{eq:RiQ}
R_i^2(Q) &=& 2 \langle(x_i-v_i(Q)t)^2\rangle-2\langle x_i-v_i(Q)t\rangle^2
\nonumber\\
&& + {1\over 2} \Big(
\langle{1\over g}
 (v_i(Q)\partial_\omega
+\partial_{Q_i})^2 g\rangle
- \langle { 1\over g } (v_i(Q)\partial_\omega
+\partial_{Q_i})g\rangle^2\Big.
\end{eqnarray}
Here the expectation value is with respect to the space-time
distribution in the source function $g(x,(\omega(Q),\vec{Q}))$,
\begin{equation}\label{eq:expec}
\left< f \right> = \frac{\int d^4x\ \ f\ \ \widetilde{g}\left(x,Q\right)}{
  \int d^4x\ \ \widetilde{g}\left(x,Q\right)},
\end{equation}
the partial derivative $\partial_\omega$ acts only on the
energy variable in the source function
and
$\partial_{Q_i}$ is a partial derivative with respect to the
spatial
$Q_i$ dependence in the source function. If the particle
distribution is in thermal equilibrium, the last two terms
in Eq. (\ref{eq:RiQ}) cancel each other. Therefore
we expect that for sources in local thermal equilibrium their
contribution will be rather small.

One point of principle is worth emphasizing at this point.  In a
classical simulation we can only get information about the source
function for on-shell momenta.  It is common to calculate a
classical, on-shell correlation function according to
\begin{eqnarray}\label{eq:cdefwig:class}
C_{cl}\left(\vec{Q},\vec{q}\right) \equiv 1 +
\frac{\displaystyle
\int d^4 x \int d^4 y\ \  g\left(x, Q+q/2\right)
 g\left(y, Q-q/2\right)
\cos\left(q (x-y)\right)}{
\displaystyle
 \int d^4 x\ \ g\left(x, Q+q/2\right) \int d^4 y\ \ g\left(y, Q-q/2\right)
}.
\end{eqnarray}
With this definition of the correlation function, the source size
evaluated from Eq.\ (\ref{eq:raddef}) has only the first two terms
in Eq.\ (\ref{eq:RiQ}), i.e.,
\begin{equation}\label{eq:radwig:class}
R_{cl,i}^2(Q) = 2 \langle(x_i-v_i(Q)t)^2
\rangle-2\langle x_i-v_i(Q)t\rangle^2.
\end{equation}
This  is  equivalent to the definition as an
r.m.s.
source size.  As discussed above, we expect the effect of the
additional terms associated with the quantum formula to be small,
but this can only be quantified with a quantum mechanical
calculation that contains the necessary off-shell information.

In the sequel we will evaluate Eq.\ (\ref{eq:cdefwig:class}) in two ways.
The first is to
replace $g\left(x,Q\right)$ by a sum over $\delta$-functions that
represents the positions and momenta of the pions at their
last interaction as  is done, e.g.,  in Ref.\ \cite{Su93}.
 The other way is to fit a smooth function
to the result of our simulation. The second choice has the
advantages
that we can demonstrate the dependence on various parameters explicitly.
In the models the source function is symmetric about the beam
axis, and we may take the spatial variables to be the
longitudinal
position $z$, the transverse
distance from the center $r_\perp$, and the angle in the
transverse plane between $\vec{r_\perp}$ and $\vec{Q}$, $\phi$.
The expressions for the three source radii are then
\begin{equation}\label{eq:rside}
R_s^2 = 2 \left<r_{\perp}^2 \sin^2\phi\right>,
\end{equation}
\begin{equation}\label{eq:rout}
R_o^2 = 2\left[
\left<\left(r_{\perp}\cos\phi - v\left(Q\right) t\right)^2\right>
-\left(\left<r_{\perp}\cos\phi - v\left(Q\right) t\right>\right)^2\right],
\end{equation}
\begin{equation}\label{eq:radlong}
R_l^2 = 2 \left\{ \left<z^2\right> - \left<z\right>^2 \right\}.
\end{equation}

%In the next section we shall parameterize the source function as
%a sum of terms in which the dependence on transverse momentum,
%transverse distance, and relative angle can be factored.

We first want to remark that the correlation
in $\phi$ will reduce the effective size of the source.  If the
source were isotropic (no $\phi$ dependence), $ \langle \sin^2
\phi \rangle = 1/2$ and the sidewards radius is just the r.m.s.
transverse radius.  If the source function contains a strong
correlation between $\hat{r}_\perp$ and $\vec{Q}$ the sideward
radius will be smaller than the actual transverse extension.
This
effect has been discussed for exploding sources or collective transverse
flow\cite{Pr84}.

For the outward radius, the expression with
a factorizable source function is
\begin{eqnarray}\label{eq:rout:fac}
R_o^2 =& & 2\left[
\left\{\left<r_{\perp}^2\right>\left<\cos^2\phi\right> -
\left<r_{\perp}\right>^2\left<\cos\phi\right>^2\right\}\right. \\
\nonumber
& + &\left. v\left(Q\right)^2 \left\{\left<t^2\right>-\left<t\right>^2\right\}
\right].
\end{eqnarray}
This effect of the $\phi$ correlation can have either sign here,
but a reduction is expected.
For a very close correlation between $\hat{r}_\perp$ and
$\vec{Q}$,
the first term reduces to a finite value, the radial dispersion
of source points, while the second term adds a dispersion in
time.  In this limit the sidewards source size goes to zero, so
one certainly expects a larger outward radius than sidewards
radius.

The longitudinal radius measures most directly the freeze-out
time of the system.  For thermal emission of a boost invariant
system at a freeze-out
proper time $\tau_f$ (see footnote\footnote{
The freeze-out time $\tau_f$ is the time of the last interaction
of a particle. This is not to be confused with the hadronization
time $\tau_h$ which is the time when a hadron emerges
from the prehadronic phase}), the longitudinal radius for particles at
rapidity $y = 0$  is given by
\begin{equation}\label{eq:13}
R_l^2 = 2 {\tau_f^2T\over m_\perp
}{K_2(m_\perp/T)\over K_1(m_\perp/T)}.
\end{equation}
In the limit $m_\perp \gg T $ this reduces to the
formula\cite{MS88} $R_l^2=2\tau_f^2T/m_\perp$ which is
sometimes used to deduce hadronization times (e.g.,
Ref.\ \cite{NA35:be}).  For realistic values of the temperature and
transverse momentum the factor $m_\perp/T$ is of the order of one
and Eq.\ (\ref{eq:13}) should be used.

\section{Numerical simulations}\label{sec:res}
For the numerical studies we used a program made from the
elements of
Boggs' cascade program \cite{Ba91}.  The
first part of the program initializes the distribution of
hadrons, which will materialize at different times.  The
parameterized
density distribution, Eq.\ (\ref{eq:1}),
 is sampled with the help of  the Metropolis
method to give a space-time distribution of hadrons having the
desired number of hadrons and a composition  specified by the parameter
$w_{\rho/\pi}$.  The materialization time of each hadron is
determined as the time when the local density from Eq.\ (\ref{eq:1}) falls
below the critical value $n_h$.  The hadrons are also given an
initial momentum, obtained by sampling the Maxwell-Boltzmann
distribution in a frame boosted to rapidity $\eta$.  The
remainder of the program is a loop stepping through time.
In each time interval, the program goes through the following
tasks:\\
\begin{enumerate}
\item
Determine which hadrons have materialized.
\item
Propagate in space according to the hadron's velocity.
\item
Collide hadrons pairwise according to the assumed cross
sections and their impact parameter. This is done using the method of
\protect{\cite{BDG88}}, with variants on how the final momentum is determined.
This will be discussed in more detail in Sec. IV D.
\item
Convert omega and rho resonances to pions according to their
time-dilated lifetimes.
\end{enumerate}

Each run of the program creates a file of the final pions
containing their momenta and the space-time coordinates of their
last interaction point.  For the studies reported here, we
typically average over 100 runs.

\subsection{Input Parameters}

As formulated, the model contains as parameters the extension of
the collision zone in spatial rapidity $\Delta \eta $, the
density at the transition to the hadronic gas $n_h$, the kinetic temperature
at the transition $T$, the proportion of vector mesons at the
transition and scattering cross sections involving vector mesons.
We first note that a number of these parameters are strongly
constrained by the empirical single-particle spectra.  In
particular, the final state rapidity distribution depends
strongly on $\Delta \eta$, weakly on $T$, and hardly at all on
any of the other parameters.  Correspondingly, the average
transverse momentum depends almost exclusively on $T$.  We may
therefore consider these parameters to be fixed by the data.
Single particle spectra for S+Au or S+Pb are not available yet,
but NA35 results for S+S at 200 GeV/u have been published\cite{Ba94}.
In the remaining calculations we take values of $\Delta \eta =
1.2$ and $T=150 $ MeV to fit the shape of the distributions from
this measurement. The quality of the fit is shown in
Fig.\ \ref{fig:sing}, where we have assumed values of the other
parameters given in Tab. \ref{tab:1}.
 We want to emphasize
again that the single-particle observables are very insensitive
to the other parameters
once the temperature and the space rapidity spread are fixed.

We determined the absolute
normalization from the number of negative particles at mid-rapidity
for  S + Au  as measured by NA35 \cite{NA35:had}.  The measured
number, $dN/dy \approx 65$, is
reproduced taking the total number of pions in
the final state to be 700.

As mentioned earlier, assumptions are required about the hadronic cross
section. The pi-pi cross section is determined from phase-shifts (cf. [21]),
but we arbitrarily assume a cross section of 20 mb for the other possible
collisions. The sensitivity to the cross section will be examined later.

The most important remaining parameter is the hadronization
density $n_h$.
For a first comparison with data we will use the
only recent publication of a correlation function, the data
of NA44 in Ref.\ \cite{Hu93}.

\subsection{Comparison to one-dimensional correlation data}
The one-dimensional correlation data of NA44 are shown in Fig. 2.
The correlation is plotted as a function of the invariant momentum
$q_{inv} = \sqrt{-q^2}$.  In this comparison, we do not use $n_h$ directly as
a parameter, instead assuming that the hadronization occurs at
a definite proper time  $\tau_h$.  We show in Fig.\ \ref{fig:a} the
model results for various $\tau_h$, and taking the proportion
of heavy mesons at $0.10$.  We applied
cuts to our pion output files that are similar to the NA44 acceptance.
It may be seen from the
curves that agreement with data can only be obtained for very
long proper times, of the order of 15 fm/c.  Using Eq.\ (\ref{eq:1}) we
find that the highest density at this time is 0.2 pions/fm$^3$,
a number which is so low that the gas would be
noninteracting at that point.
Clearly, drastic assumptions would
have to be made for the dynamics of the prehadronic phase to
obtain conversion at such extended times.

\subsection{Comparison to three-dimensional correlation data}

We now turn to the three-dimensional analyses of the source
geometry, which should in principle provide more detailed
information.  Here we shall see a less dramatic inconsistency,
but one which seems difficult to overcome without introducing
new dynamics.

The usual analyses of source geometry assume Gaussian source
shapes, which actually do not describe the source shape well.
For that reason the extracted radii bear little relation to the
definition, Eq.\ (\ref{eq:raddef}).  We illustrate this with a typical
calculation, with parameters and results shown in Tab.\ \ref{tab:1}.
This parameter set is fit to the inclusive spectra as well as the
Gaussian longitudinal radius from NA35\cite{NA35:be}.
The favored hadronization density, $n_h = 1.0$ pions/fm$^3$, corresponds
to an average hadronization proper time, $\tau_h = 2.3$ fm with a standard
deviation $\sigma = 1.5 $ fm.
The mean hadronization time
is large compared to theoretical estimates
of the equilibration time of the prehadronic phase\cite{MQ86,BM87}, supporting
our assumption of local thermal equilibrium.
% This seems to be large enough
%to  justify our assumption of local thermal equilibrium.
The spread in proper hadronization time is too big to allow for the
simplification of a hadronization at a particular proper time instead
of the more physical assumption of a hadronization density.
The  first row in Tab.\ \ref{tab:1} shows the
extracted source radii from the operational definition,
Eq.\ (\ref{eq:raddef}). This equation can only be used
directly if we know the correlation function with a high
resolution for small relative momenta. To circumvent the problem
with low statistics we fit a smooth function to the distribution
of last interaction points from the simulation. This is shown
in detail in App.~\ref{sec:par}.

 These numbers are quite large due to the decay of
the omega meson at large distances from the reaction zone.
Removing particles that decay after $\tau=20$ fm/c gives the
numbers in the second row.  We see that they are considerably
smaller.  Next we fit to a Gaussian
parameterization of the correlation function,
\begin{equation}\label{eq:G}
C_f = 1 + \lambda \exp
(-{1\over2}(q_l^2R_l^2+q_s^2R_s^2+q_o^2R_o^2)).
\end{equation}
Numerically, we make the fit following the same procedure
used by the experimentalists.  Namely, we  bin the
three-dimensional correlation function (bin width
10 MeV/c), and perform a four-parameter least squares fit.  We
take pions in the interval 100 MeV/c $< p_\perp < 300 $ MeV/c
and $\left|y\right| < 1$.  The
results are shown in the third row.  These radii are
within 1 fm of the true source radii with late decays
excluded.
The table also shows the Gaussian fit with the late-decaying
particles
excluded. This gives  practically identical radii, although the
value of $\lambda$ is somewhat changed.
Thus,
the 3D Gaussian fit is insensitive to the contribution of the
long-lived resonances.  One last point to note is that the
difference between outward and sideward source radii is
rather suppressed in the 3D Gaussian fit.

The experimental results for reactions at 200 GeV/u for various
projectile-target systems have been published in conference
proceedings\cite{NA35:be,NA44:be,NA35:be91}. There are no high
precision
data from two different collaborations
for the same target-projectile combination, but the S+Pb system
studied by NA44 should be very similar to the S+Au system studied
by NA35.  However, the numbers for the source radii quoted by
these groups are quite different.  In Tab.\ \ref{tab:1}, we have extracted
numbers for the the NA35 experiment from their Figs.\ 2,3 and 4
in \cite{NA35:be}, estimating an average size for the momentum
range $100$ MeV/c $<p_\perp<300$ MeV/c.  For the NA44 numbers,
we quote  the $\pi^+$ and $\pi^-$ radii for the S+Pb system
in Tab.\ \ref{tab:2} of Ref.\ \cite{NA44:be}\footnote{The NA44  numbers are
changed by a
factor of $\sqrt{2}$ to correspond to the definition
Eq.\ (\protect{\ref{eq:raddef}}).}.

The baseline parameter set for our model is chosen to fit the
NA35 longitudinal source radius\footnote{The use of
Eq.\ (\protect{\ref{eq:13}})
to translate this radius into a freeze-out time leads to $\tau_f = 2.5 $ fm/c,
which is much smaller than the actual average last interaction time.
The Eq.\ (\protect{\ref{eq:13}}) cannot be used because it was derived
assuming boost invariance. The temperature is also
much lower when the particles freeze-out than when they hadronize.}
 as well as the single-particle
distribution.  Note that neither the outward nor the sidewards
radii are fit here.

We next examine the effects of varying the parameters.  A number
of cases are shown in Tab. \ref{tab:2}, compared with the results of the
baseline parameter set.  The second line shows the effect of
increasing the number of primordial vector mesons in the hadronic
phase by 25\%.  There is virtually no effect on the radii, but
the normalization $\lambda$ is lowered.  We next examine the
effect of a different cross section in the hadronic phase.  The
baseline parameters were modified by doubling all the hadronic
cross sections for the entries in the third line of the table.
Again, the effect on the radii is very slight.  Finally, the
fourth and fifth lines show the effect of changing the
hadronization density by $+50\%$ and $-60\%$, respectively.
It may be seen that this has a strong effect on the longitudinal
radius.  Thus, the longitudinal radius gives a good measure of
the hadronization density.  There is also a significant effect on
the outward radius.  This is because the outward radius depends
on the time distribution of the emission of hadrons, which is
broader when the conversion is a lower density.  The sidewards
radius is hardly affected by the hadronization density, but this
is a consequence of our basic assumption that there is
no sidewards expansion in the prehadronic phase.

The parameter changes we have discussed do not affect the
one-particle observables significantly.  The two remaining
parameters, $T$ and $\Delta \eta$, cannot be changed without
disturbing the agreement there.

\subsection{Dependence on cross section assumptions}

We now examine the sensitivity of the predicted source radii
to the assumptions about the two-body collisions. The base model uses
the collision algorithm directly from Ref. \cite{Ba91}. In the code
the final state momenta in such a collision are in the reaction
plane and distributed uniformly in scattering angle $\theta$.
The momentum transfer is always chosen to repel one particle
from the other one. We call these assumptions the "repulsive, uniform
in $\theta$" model. Table \ref{tab:drei} shows the base model
source radii once more, together with results when these assumptions
are altered.

The first change we make is to double all cross sections, including
the pi-pi cross sections. It may be seen from the second line in the
table that this produces a completely negligible effect on the radii.
Within the context of the repulsive algorithm, we can make
changes to the angular distribution which are also unimportant
for the observables. The third line shows the results when the angular
distribution is taken to be uniform in $\cos \theta$, which corresponds
to isotropic angular distributions when averaged over many collisions,
i.e., directions of the reaction plane. For the fourth entry
we used the differential cross section associated with the empirical
phase shifts in the pi-pi scattering cross section. All of these
changes have negligible effect on the extracted source sizes.

Recently it has been found that the direction of the momentum
transfer can effect observables. Ref. \cite{K} shows that flow is
effected at Bevalac energies, and Ref. \cite{D} argues that the
assumptions here can effectively introduce a pressure into the
equation of state.

We next present the results keeping the scattering in plane, but
reversing the momentum transfer from repulsive to attractive.
The effects on the longitudinal and outward radii are surprisingly
large, as shown in the fourth entry of  Table \ref{tab:drei}.

Both these radii increase by $\approx 50 \%$. Qualitatively, an increase
is expected because the attractive algorithm corresponds to a negative
pressure contribution which slows down the expansion and keeps the
system interacting to larger radii. We feel that the attractive
algorithm is an unreasonable extreme, a more moderate assumption
is to distribute the  final momenta uniformly in solid angle. This
is shown as the last entry in Table \ref{tab:3}. The increase
in $R_l$ and $R_o$ is then $\approx 15 \%  $.  The results of the
"repulsive, uniform in $\theta$" model are reproduced when we
increase of the hadronization density by $25 \%$.

It should be mentioned that the single particle distributions we used
for the determination of the model parameters are unaffected by these
changes in the cross sections.

\subsection{Transverse Momentum}
Recently NA35 has measured the dependence of the source size on
transverse momentum\cite{NA35:be}, finding that the size shrinks
as the transverse momentum is increased.
This effect was predicted by Pratt \cite{Pr86} and is due to the
collective flow, which originates in our model in the rescattering
of the hadronized pions.
 Without this rescattering, our model would have no
correlation between position and momentum and the source size
would reflect the spatial extension of the entire source.  The
correlation between transverse momentum and position must vanish
as the momentum goes to zero, so the low-momentum source size
is not affected by this correlation.  On the other hand, at
finite momentum the correlation leads to a reduction of the
measured transverse size of the system. In Fig.\ \ref{fig:pp} (a)
we compare the results from our simulation to the ones obtained by
NA35 \cite{NA35:be}. As before our results turn out to be much smaller
than the experimental findings. However,
the momentum dependence of the size
agrees with the trend of the data.

In Fig.\ \ref{fig:pp} (b) we show the difference between outward and sideward
radius as a function of the transverse momentum. Here the experimental results
\cite{NA35:be} and the results from our simulation are essentially
compatible with the difference being zero. Analogous to
the results summarized in Tab. \ref{tab:1} this is, at least in our
model, an artifact of the 3-D Gaussian fitting. Therefore it is probably
safe to conclude that there is no {\it very} long-lived source. However,
one is {\it not} forced to conclude that the particles are emitted
at the same time, either.

\section{Conclusions}\label{sec:con}
In our conclusions, we first emphasize that the source radii one extracts
from pi-pi correlations depend very much on how the analysis is made.
A proper definition of the radius, such as in Eq.\ (\ref{eq:raddef}),
gives large radii
with substantial differences in the three directions. These actual radii
have a large component due to the omega meson decay, however,
and they could only be measured by having momentum accuracies to much
smaller than 10 MeV/c. The usual technique for extracting source radii,
fitting a 3-D Gaussian as in Eq. (15), turns out to be insensitive
to the resonance decays. However, there is also a much reduced
discrimination between the different directions. In particular the
actual outward radius is twice as large as the sideward one, but
this is reduced to a 10 \% difference in the 3-D Gaussian fit.

We have shown that a purely hadronic model, consisting of light mesons
interacting by two-particle scattering is inconsistent with the measured
longitudinal radii.

Quantitatively we find that the hadronic gas picture needs to be replaced
by other dynamics when the energy density is higher than about
0.8-1.0 GeV/fm$^3$.
The local temperature at the conversion point is constrained by the
transverse momentum spectrum to be about 150 MeV.  Theoretically,
one no longer expects a first-order phase transition between the
hadronic phase and a
high-temperature quark-gluon plasma phase\cite{Ch90}, as was
first
suggested. However, a
remnant of that transition may persist producing a large increase of
energy and entropy density over a small interval of temperature.
Estimates for this temperature are in the range of 150
MeV\cite{Ch92}.  The energy density of a gas of gluons and quarks
of two flavors is of the order
of 1 GeV/fm$^3$ at this temperature.  Just below the transition
temperature, the energy density would be that of  a gas of
pions and the light vector mesons, which is  0.1
GeV/fm$^3$ for $T=150$ MeV.  By fitting the density of conversion
to the experimental longitudinal radius, we are led to a hadronic
gas that is far out of chemical equilibrium.  Further work is
necessary before conclusions can be drawn about the nature of
this
transition.  However, it is intriguing that the energy density of
a quark-gluon plasma is so close to the results of the model.

We also found that the predicted sidewards and outward radii were
consistently small compared to experiment.

 A similar conclusion was reached in \cite{Wel94}, for O+Au reactions,
using a model similar to that employed here. Also, a smaller
sidewards size than observed was found\cite{We94a} in Werner's
string model\cite{We93}.  In this model, the prehadronic phase is
described by strings which decay into prehadronic clusters.
Like in our models, there is no transverse dynamics in the
prehadronic phase.

The small predicted sidewards radius is particularly difficult to
reconcile, because the obvious remedy, an adiabatic transverse expansion,
appears to have  little effect on this observable\cite{We92,Sc92}.
Thus a transverse expansion with some nontrivial dynamics
seems to be required.  This would be indirect evidence of the
pressure in that phase; if there were a phase transition the
pressure should be quite low compared to the energy density.
Any transverse expansion in the prehadronic phase would lower
the required density at the conversion point. Thus our number
0.8-1.0 GeV/fm$^3$ should be considered an upper limit.
More work obviously needs to be done to explore the bounds
on the equation of state provided by the data.

Finally we mention that the more inclusive correlation observable,
given by the invariant momentum distribution, requires an unphysically
large hadronization time, because the average size must be fit by adjusting
only one of the three independent dimensions.

\acknowledgements
We would like to thank
J.~Cramer,
D.~Ferenc,
T.~Humanic,
B.~Jacak,
R.~Morse,
G.~Roland,
R.~Stock,
L.~Teitelbaum
and L.~Y.~Yan
for helpful conversations about the experimental measurements.
We would also like to thank P.~Danielewicz, G.~Welke and
H.~Schulz for discussions leading to this study.  This work has
been supported in parts by the Department of Energy
under Grant DE-FG06-90ER40561
(G.B., M.H.), and
by the Alexander von Humboldt-Stiftung
(Feodor-Lynen Program) (M.H.).

\appendix
\section{Source parameterization}\label{sec:par}
We display in this appendix a parameterization of single-pion
source function obtained with our baseline model.  This
parameterization is useful for determining the actual source
radii from Eq. (\ref{eq:raddef}).  It is also useful to see the influence of
the
long-lived resonances, which appear in the parameterization as
an exponentially decaying component to the source distribution.

The distribution of the
emission points of the pions can be written as a function of proper
time $\tau$,  space rapidity $\eta$, the transverse distance
to the beam axis $r_{\perp}$ and the angle between the transverse component
of the position and the transverse component of the momentum
$\phi \equiv \angle \left(\vec{p_\perp},\vec{r_\perp}\right)$. These
are still too many variable to make a general fit, therefore we
assume that we can separate the emission into several intervals
in proper time. For each of these intervals we assume, that the emission
function can be factorized in the other variables, i.e.,
\begin{equation}\label{eq:source}
g\left(\tau, \eta, r_{\perp}, \phi \right) =
g^{\tau}_i\left(\tau\right) g^{\eta}_i\left(\eta\right)
g^{r}_i\left(r_\perp\right)
g^{\phi}_i \left(\phi\right),\ \  \tau_i<\tau<\tau_{i+1}.
\end{equation}
In Fig. \ref{fig:tau} we show the proper time distribution of the pion
emission, considering  100 runs taking pions
with rapidity $\left|y\right| < 1$ and transverse momentum in the range
 100 MeV/c $ < p_{\perp} <$ 300 MeV/c. This distribution can be
conveniently
separated into  three intervals:\\
1. 0 fm/c $ < \tau < $ 10 fm/c, 2. 5 fm/c $<\tau < $ 10. fm/c
and 3. 10 fm/c $ < \tau$. The boundaries of these regions are indicated
by arrows in the figure.  The distribution is well fit by a Gaussian in the
first interval
and exponentials in the later intervals, as shown by the lines in
Fig.\ \ref{fig:tau}.
The $r_{\perp}$ distribution is fit by Gaussians in the first and second
interval and by an exponential in the third one.
 We also found that the $\eta$  distribution is
reasonably fit by Gaussians. To determine the $\eta$ distribution we
transformed  each pion into its rest system.

For the $\phi$ distribution, we need information about the first
two coefficients of a Fourier expansion to calculate the
outward and sidewards radii.  We thus use a parameterization of
the form
\begin{equation}\label{eq:four}
g^\phi = 1 + b \cos\left(\phi\right) + c \cos\left(2 \phi\right)
\end{equation}
In Fig. \ref{fig:phi} we show how well this parameterization compares with
the output of the simulation program.  Note that the correlation
is quite strong between transverse radius and momentum.

The  parameterization of the
dependence on proper time and the other variables is summarized in
Tab.\ \ref{tab:3}.

In Fig. \ref{fig:comp} we show the results for the outward and sideward
correlation function. Thereby we assumed all pion pairs to to have an average
momentum of 200 MeV/c. The crosses are obtained by using the
results from our simulation directly, i.e., replacing the integral
over the source function by a sum over delta functions. The dashed line
shows the appropriate component of the three-dimensional Gaussian fit,
with the parameters as given in Tab.\ \ref{tab:1}. The full line
is the result using the parameterization of the source function.
Note that the sum over delta functions and the Gaussian fit lead
to a ``coherence'' parameter $\lambda < 1$ whereas the correlation
function extracted from the parameterization goes to 2 for small $q$.
The deviation of $\lambda$ from one in the former case is entirely
due to the finite binning and the limitations of a Gaussian fit, respectively.
For relative momenta $q\geq 100$ MeV/c our parameterization
gets unrealistic as we have assumed that the position-momentum correlation
is independent of $r_\perp$. This assumption is justified for the bulk part
of the particles. For particles that are emitted at $r_\perp$ small the
emission has to be isotropic and therefore $g^\phi$ had to go to 1 in this
limit. For the extraction of the radii that are determined by the low $q$
behavior this has no consequences.

\begin{table}
\caption{Source radii for S + heavy target.  Parameter values
are: $n_h=1.0$  pions/fm$^3$, $w_{\rho/\pi}=0.16 $, $T=150$ MeV,
$\Delta\eta=
1.2$.
}
\begin{tabular}{|c|c|c|c|c|}
\hline
 & $R_l$ & $R_o$ & $R_s$& $\lambda
 $ \\
 & (fm)  & (fm) & (fm)&
see footnote\footnote{
Our simulation does not include the effect of $\eta$ meson decays
on the correlation function, which will only lead to a change
of the intercept. Therefore $\lambda$ is listed for the sake of completeness
only.}
 \\ \hline
Eq. (5)  & 12.&15. &7.5 & 1 \\
 $ \tau < 20 $fm/c & 4.3  &4.5 & 2.9& 1\\
fit to Eq. (15) &4.0 & 3.6 & 3.2 & 0.82  \\
as above, $\tau < $ 20 fm/c&4.0 &3.6 & 3.2 & 0.94 \\ \hline
NA35  & 4  &  4+ & 4+&  \\
NA44  & 6.2-6.9  &5.1-5.8 &5.2-6.2 & 0.6-0.7 \\    \hline
\end{tabular}
\label{tab:1}
\end{table}

\begin{table}
\caption{Sensitivity of source radii to model parameters.}
\begin{tabular}{|c|c|c|c|c|}
\hline
 & $R_l$ & $R_o$ & $R_s$& $\lambda$ \\
 & (fm)  & (fm) & (fm)& \\ \hline
base  & 4.0 & 3.6 & 3.2 & 0.82 \\
$w_{\rho/\pi}=0.21$&4.0& 3.7& 3.2& 0.78\\
$n_h=1.50 $&3.2&3.4&3.2&0.86\\
$n_h=0.40 $&8.2&5.6&3.0&0.77\\
\end{tabular}
\label{tab:2}
\end{table}

\begin{table}
\caption{Sensitivity of source radii to two-body collision assumptions.}
 \begin{tabular}{|c|c|c|c|c|}
\hline
 & $R_l$ & $R_o$ & $R_s$& $\lambda$ \\
 & (fm)  & (fm) & (fm)& \\ \hline
base: repulsive, uniform in $\theta$& 4.0&3.6&3.2&0.82\\
base $\times 2$& 4.0&3.6&3.4& 0.88\\
 repulsive, uniform in $\cos \theta$&4.0&3.6&3.2&0.83\\
repulsive, s+p& 3.9 & 3.6 & 3.1& 0.82\\
attractive, uniform in $\theta$& 6.4& 6.8 & 3.2 & 0.74\\
isotropic & 4.6 & 4.3& 3.2 & 0.79
\end{tabular}
\label{tab:drei}
\end{table}

\widetext
\begin{table}
\caption{Functions and parameters for the parameterization of the
source function. Units are fermi for lengths and fermi/c for times.}
\begin{tabular}{|c|c|c|l|}
\hline
i & 1 & 2& 3\\ \hline\hline
$n_i$& 128 & 15 & 11\\ \hline\hline
$g^{\tau}_i$&$a_1^{\tau}\exp\left(\frac{
\left(\tau-b_1^{\tau}\right)^2}{2 c_1^{\tau^2}}\right)$&
\multicolumn{2}{c|}{$a_i^{\tau} \exp\left(-\tau/b_i^{\tau}\right)$}
\\ \hline &$a_1^{\tau} = 0.18, b_1^{\tau}=5.0, c_1^\tau=2.2$ &
$a_2^{\tau}=1.3, b_2^{\tau}=5.6$
& $a_3^{\tau} = 0.11, b_4^{\tau} = 22.$\\
\hline\hline
$g^r_i$&
\multicolumn{2}{c|}{
$ a^r_i r_{\perp} \exp\left(-\frac{\left(r_{\perp}-b_i^r\right)^2}{
 2 c_i^{r 2}}\right)$}&
$a^r_3 r_{\perp} \exp\left(-r_{\perp}/b_3^r\right)$ \\
\hline
&$ a_1^r = 0.20, b_1^r = 0.59, c_1^r = 1.84$&
$a^r_2 = 0.043, b^r_2 = 0.010, c^r_2 = 4.8 $&
$a^r_3 = 0.010, b^r_3 = 9.9$
\\ \hline \hline
$g^{\eta}_i$&\multicolumn{3}{c|}{
 $\left(\sqrt{2\pi}\Delta \eta\right)^{-1} \exp\left(
-\frac{\eta^2}{2 \Delta\eta_i^2}\right)$}\\\hline
 & $\Delta \eta_1 = 0.47$&$\Delta\eta_2 = 0.31$&$\Delta\eta_3=0.57$
\\\hline\hline
$g_i^{\phi}$&\multicolumn{3}{c|}{
$1+b_i^{\phi} \cos\left(\phi\right) + c^{\phi}_i \cos\left(2 \phi\right)$}\\
\hline
&$ b_1^{\phi} = 1.2, c_1^{\phi} = 0.41$
&$ b_2^{\phi} = 1.4, c_2^{\phi} = 0.95$
&$ b_3^{\phi} = 0.95, c_3^{\phi} = 0.13$
\end{tabular}
\label{tab:3}
\end{table}
\narrowtext

\begin{figure}
\caption{The single-particle distributions $dN/dy$ and $dN/p_{T}
d p_{T}$
from \protect{\cite{Ba94}}
(full squares) for $S+S$ and our simulation (full line) using the base set of
parameters.
%The experimental data are the ones obtained with the VETO trigger
%as described in \protect{\cite{BA94}}.
We constructed the open
squares in (a) by assuming symmetry around CM rapidity $y = 3$. The data
in (b) are for the laboratory rapidity window $2<y_L<3$. The
normalization of our results in both figures is arbitrary.}
\label{fig:sing}
\end{figure}

\begin{figure}
\caption{Correlation function as a function of the invariant momentum for
different values of the hadronization time $\tau_h$. We used our base set
of parameters except for the vector meson to pion ratio that was chosen
to be $w_{\rho/\pi} = 0.10$.
The crosses show
the NA44 results from Ref. \protect{\cite{Hu93}}.}
\label{fig:a}
\end{figure}

\begin{figure}
\caption{The sideward radius (a) and the difference between sideward
and outward radius (b) as a function of the transverse momentum. The
dots are NA35 data\protect{\cite{NA35:be}} for a laboratory rapidity
$3.6 < y_L < 4.6$. Our results (squares) are obtained by a
3D-Gaussian fit using the pions with a CM rapidity $0.6<y<1.6$. }
\label{fig:pp}
\end{figure}

\begin{figure}
\caption{The number of pions emitted per unit proper time interval
as a function of proper time. The  arrows  indicate the boundaries
of the intervals
we chose for our parameterization. The full lines show our parameterization.}
\label{fig:tau}
\end{figure}

\begin{figure}
\caption{Correlation between position and momentum of emitted
pions, for the group emitted before 10 fm/c.  The dots show the
probability distribution $g_1^\phi$ from the model with the base
parameter set, and the line shows the fit with Eq.\ (\protect{\ref{eq:four}}).}
\label{fig:phi}
\end{figure}

\begin{figure}
\caption{ Comparison between different methods to extract a correlation
function from our simulation for the outward (a) and
sideward (b) projection. The crosses are the results obtained by
replacing the source function by a sum over delta functions given
directly by the final distribution of momenta and last interaction
points. The dashed line shows the result of a 3D Gaussian fit to these
points. The full line is the correlation function calculated using
the parameterization Eq.\ (\protect{\ref{eq:source}}). Note  that the result
from the parameterization goes to two for $q_{out/side} = 0$ whereas
the other methods show a smaller intercept due to finite binning
over ranges of  $y$ and $p_T$.}
\label{fig:comp}
\end{figure}

\end{document}